\documentclass[twocolumn,aps,prd,%nofootinbib,
superscriptaddress,tightenlines]{revtex4-1}

\usepackage{amsmath, amsfonts, amsthm, amssymb, graphicx, color,hyperref}
\usepackage{subfigure}

\allowdisplaybreaks[3]
\def \bal#1\eal  {\begin{align} #1 \end{align}}
\newcommand{\be} {\begin{equation}}
\newcommand{\ee} {\end{equation}}

\def\ba{\begin{eqnarray}}
\def\ea{\end{eqnarray}}
\newcommand{\ud} {\mathrm{d}}

\newcommand{\pd} {\partial}

\newcommand{\mc} {\mathcal}
\newcommand{\tld}{\tilde}
\newcommand{\ai}{{\alpha}}
\newcommand{\bi}{{\beta}}

\newcommand{\ri}{{\rho}}
\newcommand{\si}{{\sigma}}
\newcommand{\li}{{\lambda}}

\newcommand{\ep}{{\epsilon}}
\newcommand{\mpl}{M_P}
\def\mupn{^\mu_{\, \nu}}

\begin{document}

\title{Can the graviton have a large mass near black holes?}

\author{Jun Zhang}
\email{jun34@yorku.ca}
\affiliation{Department of Physics and Astronomy, York University, Toronto, Ontario, M3J 1P3, Canada}
\affiliation{Perimeter Institute for Theoretical Physics, Waterloo, Ontario N2L 2Y5, Canada}

\author{Shuang-Yong Zhou}
\email{zhoushy@ustc.edu.cn}
\affiliation{Interdisciplinary Center for Theoretical Study, University of Science and Technology of China, Hefei, Anhui 230026, China}
\affiliation{Theoretical Physics, Blackett Laboratory, Imperial College, London, SW7 2AZ, U.K. }

\date{\today}

\begin{abstract}

The mass of the graviton, if nonzero, is usually considered to be very small, {\it e.g.} of the Hubble scale, from several observational constraints. In this paper, we propose a gravity model where the graviton mass is very small in the usual weak gravity environments, below all the current graviton mass bounds, but becomes much larger in the strong gravity regime such as a black hole's vicinity. For black holes in this model, significant deviations from general relativity emerge very close to the black hole horizon and alter the black hole quasi-normal modes, which can be extracted from the ringdown waveform of black hole binary mergers. Also, the enhancement of the graviton mass near the horizon can result in echoes in the late time ringdown, which can be verified in the upcoming gravitational wave observations of higher sensitivity. 

\end{abstract}

\maketitle

The recent detection of gravitational waves~(GWs) by the advanced Laser Interferometer Gravitational Wave Observatory (LIGO) \cite{Abbott:2016blz} has ushered in the era of GW astronomy, opening up a new window to probe black holes as well as to test general relativity (GR) and its alternatives in the strong gravity regime. 

A natural alternative to GR is to let the graviton mass be nonzero.  Indeed, the observation of the late time cosmic acceleration has triggered a surge of interest in massive gravity where the graviton has a Hubble scale mass, especially after the discovery of de Rham-Gabadadze-Tolley (dRGT) theory \cite{deRham:2010kj, deRham:2010ik,Hassan:2011hr, Hassan:2011ea}, which overcomes the longstanding theoretical pathology of the Boulware-Deser (BD) ghost \cite{Boulware:1973my}. The original dRGT model, however, does not accommodate stable homogeneous and isotropic cosmological solutions \cite{DeFelice:2012mx}, nor does it admit healthy static and spherically symmetric black holes \cite{Volkov:2014ooa, Berezhiani:2011mt}. These problems can be circumvented in mass-varying massive gravity (MVMG) \cite{Huang:2012pe}, in which the dRGT graviton mass is promoted to a function depending on a scalar field and thus can vary with the environment.

In this paper, we investigate how much the graviton mass can vary with the environment without running into trouble with observations. We will construct a massive gravity model where the graviton mass is very small ({\it e.g.}, of Hubble scale\;$10^{-33}\text{eV}$) in weak gravity environments like the solar system or cosmological settings, but becomes much larger ({\it e.g.}, $10^{-10}$eV) near the horizon of astronomical black holes, yet still phenomenologically viable. Novel features arise in this model and can be examined in the upcoming GW observations of higher sensitivity. Especially, the enhancement of the graviton mass near the black hole horizon can result in late time ``echoes'' in the ringdown waveform of black hole mergers, similar to those induced by exotic matters or features near the horizon \cite{Cardoso:2016rao, Cardoso:2016oxy}. Do echoes exist in the LIGO data? How can they be further extracted if exist? These questions have become an active subject of recent discussions \cite{Abedi:2016hgu, Ashton:2016xff, Abedi:2017isz, Price:2017cjr,Cardoso:2017njb,Mark:2017dnq}.

Consider a massive gravity model given by the action
\begin{equation}
\label{mvmg}
S=\mpl^{2}\!\! \int \!\!\ud^{4} x \sqrt{-g}
\left[ \frac{R}{2} \! + \! V(\si)\, {\cal U}\! -\! \frac12 (\pd \sigma )^2 \! - \! W(\si) \right], 
\end{equation}
where ${\cal U} =  U_2 + \alpha_3 U_3 + \alpha_4 U_4$ is the dRGT graviton potential and $U_i\equiv \mc{K}^{\mu_1}_{[\mu_1}... \mc{K}^{\mu_i}_{\mu_i]}$, $\mc{K}\mupn\equiv \delta^\mu_\nu - \sqrt{g^{-1} \eta}\big|^\mu_\nu$, with $g^{-1}=(g^{\mu\nu})$ and $\eta=(\eta_{\mu\nu})$. Here the anti-symmetrization is defined with weight one ({\it e.g.}, $A_{[\mu\nu]}=(A_{\mu\nu}-A_{\nu\mu})/2$). The graviton mass\,\footnote{The precise definition of a particle mass in curved space is a subtle issue. Here we simply take the size of coefficients of the quadratic perturbative potential as an effective measure.} is regulated by the environment field $\si$. We will focus on the simple model with
\begin{equation}
V(\si) = m_0^2 + m^2 \sigma^{4},~~~~ W(\si)=\frac{1}{2}m_{\si}^2\si^2+\lambda_\si \si^6   ,
\end{equation}
where $m_0$ is assumed to be of the cosmological scale, potentially accounting for the late time cosmic acceleration but playing little role for astronomical black holes. In contrast, a typical value of $m$ is taken to be around the inverse of the Schwarzschild radius of a stellar black hole.

In the weak gravity regime, the environment field $\si$ sits around zero due to the absence of direct decoupling between $\si$ and the matter fields, thus the model essentially reduces to the original dRGT model. Near black holes, $\si$ can grow nontrivially. To obtain the black hole solutions, we start with the most general static and spherically symmetric ansatz:
\begin{align}
\ud s^2  &= - a(\tld{r}) \ud \tld{t}^2 + 2 b(\tld{r}) \ud  \tld{t} \ud \tld{r} + c(\tld{r}) \ud \tld{r}^2 + d(\tld{r}) \ud \Omega^2,  \nonumber
\\
\ud & s_\eta^2  =   - \ud \tld{t}^2  + \ud \tld{r}^2 + \tld{r}^2 \ud \Omega^2 \quad {\rm and}\quad \si =\si(\tld{r}).
\end{align}
There are two branches of solutions \cite{Tolley:2015ywa}. The diagonal branch ($b=0$) leads to GR-like black holes that are plagued by various pathologies (e.g.~\cite{Berezhiani:2011mt}) and is thus not of interest here. The non-diagonal branch ($b\neq 0$) leads to interesting hairy black holes, which requires $\beta k_3^2 + 2\alpha k_3 + 1 = 0$ with $\ai \equiv 1+ \alpha_3$ and $\bi \equiv \ai_3+\ai_4$. $k_3 \equiv 1-\sqrt{\tld{r}^2/d(\tld{r})}$ is the eigenvalue of the matrix $\mc{K}$ that has algebraic multiplicity 2. 

For the non-diagonal branch, the equations of motion reduce to a system of ordinary differential equations for the variables $a(\tld{r})$, $e(\tld{r}) \equiv ac+b^2$  and  $\si(\tld{r})$. For boundary conditions, one has $a(\tld{r}_h)=0$ at the black hole horizon ($\tld{r}=\tld{r}_h$) due to the fact that the horizon of a static and spherically symmetric black hole is necessarily a Killing horizon of $\pd_{\tld{t}}$. The requirement that the horizon is free of physical singularities imposes another boundary condition on $\si'(\tld{r})$ at $\tld{r}_h$. Also, we are interested in flat asymptotics \footnote{We are working in the limit where $m_0$ vanishes, as for astrophysical black holes $m_0$ is negligibly small.}, which imposes the other two boundary conditions that $\si \to 0$ and $b \to 0$ at large $\tld{r}$. Generically, the solution of such a system can be obtained numerically via a 2D shooting procedure. The hairy black hole obtained in this way is {\it unique} for a given (Arnowitt-Deser-Misner) mass $M_{\rm bh}$.

Our numerical integration starts at a small distance away from the horizon and goes to a sufficiently large $\tilde{r}$. Viewed as an initial value problem with ``time'' variable $\tilde{r}$, the Cauchy data near the horizon are $e$ and $\si$, which should be tuned so that the two boundary conditions at large $\tld{r}$ are satisfied. For simplicity, we choose the model parameters $\ai_3=-2/k_3$, $\ai_4=3/k_3^2$, which allows us to shoot $e$ and $\si$ independently. We have tried a few other choices of these parameters and found the solutions do not differ qualitatively.

\begin{figure}[tbp]
\centering 
\includegraphics[height=0.22\textwidth]{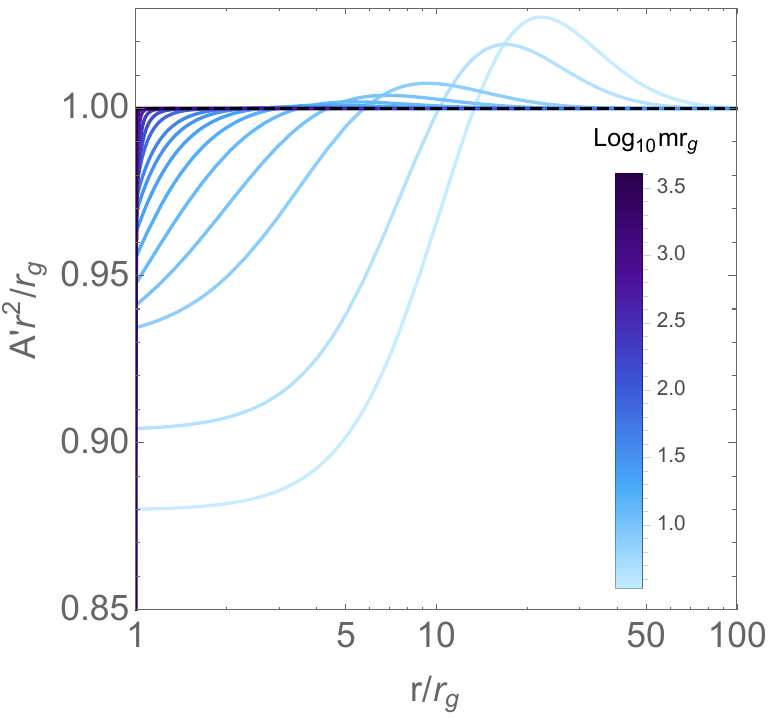} 
\includegraphics[height=0.22\textwidth]{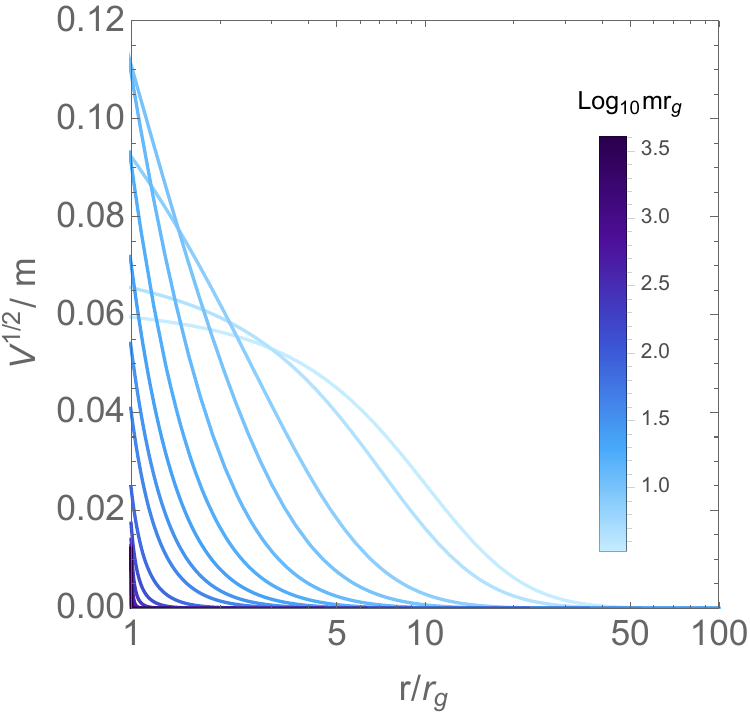} 
\caption{Numerical black hole solutions. The parameters chosen are $m_\si=10^{-2}m$, $k_3 = -1/10$, $\ai_3=-2/k_3$, $\ai_4=3/k_3^2$ and $\lambda_\si=0$. $A$ is the $tt$ metric component, $V^{1/2}$ is the graviton mass and $r_g=2GM_{\rm bh}$.}
\label{fig:sol}
\end{figure}

\begin{figure}[tbp]
\centering 
\includegraphics[width=0.23\textwidth]{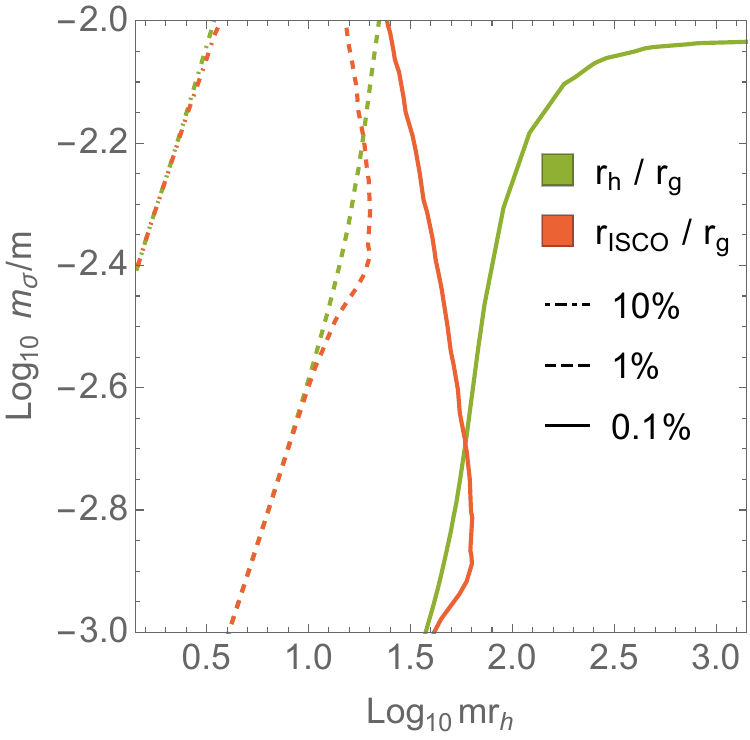}
\includegraphics[width=0.23\textwidth]{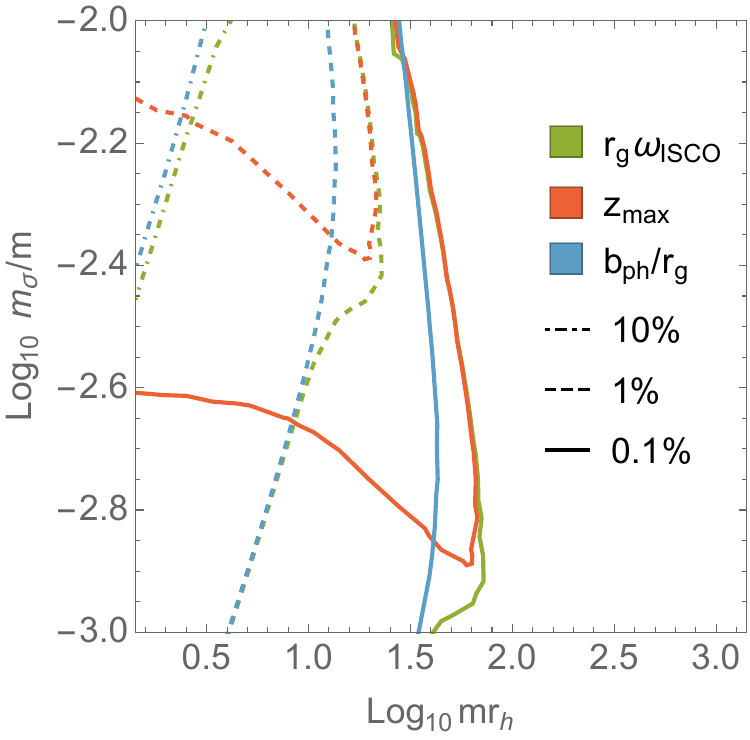} 
\caption{Deviations from GR for geodesics near black holes.  We choose $k_3 = -1/10$ and $\lambda_\si=0$. The contours depict fractional deviations (0.1\%, 1\% and 10\%) of the relevant quantities from their GR counterparts.}
\label{fig:para}
\end{figure}

To present the black hole solutions, it is instructive to rescale the time and radial coordinate $t=\sqrt{a_\infty}\tld{t}$, $r=\sqrt{c_\infty} \tld{r}$ so that the dynamical metric $g_{\mu\nu}$ approaches the standard Minkowski metric $\eta_{\mu\nu}$. In the new coordinates, the $tt$, $tr$ and $rr$ metric components are respectively 
\begin{equation}
A =  \frac{a}{a_\infty}, ~~B = \frac{b}{\sqrt{a_\infty c_\infty}}, ~~{\rm and}~~C= \frac{c }{c_\infty} \, ,
\end{equation}
where $a_\infty=4(1-k_3)^4/(k_3^3-6k_3^2+6k_3-2)^2$ and $c_\infty=1/(1-k_3)^2$. See Fig.~\ref{fig:sol} for a few fiducial black hole solutions. The hairy solutions deviate from Schwarzschild spacetime near the horizon, and settle down to it quickly as $r$ increases. From the $1/r$ fall-off behavior of the $tt$ metric component, we can infer the gravitational radius $r_g = 2GM_{\rm bh}$. As shown in the left panel of Fig.~\ref{fig:para}, the gravitational radius is generically greater than the horizon radius. The graviton mass $\sqrt{V}$ decreases exponentially to zero with $r$  (since we have neglected $m_0^2$), so typically within a couple of $r_h$ away from the horizon, the metric essentially reduces to the Schwarzschild metric, implying that all the current weak field GR tests can be easily passed as we shall see later. In the vicinity of the horizon,  however, $\sqrt{V}$ can be of the order of $1/r_h$ or even larger. For a given $m$, the region where solutions significantly deviates from GR decreases as the black hole mass $M_{\rm bh}$ increases. 
%%%%%%%%%%%%%%%%%
%%%%%%%%%%%%%%%%%

Before discussing the observational aspects of these solutions, we first make a couple of theoretical remarks. On the asymptotic vacuum of these black holes, the $V\mc{U}$ term contributes a negative $\si^{4}$ term to the $\si$ potential. The (positive) $\li_\si$ term of $W$ is needed precisely to render the $\si$ potential globally bounded from below, but the black hole solutions are insensitive to the choice of this nonlinear term in $W$, given the fact that $\si(r) \ll1$ for these solutions. So $\si = 0$  might be a local minimum and is meta-stable quantum mechanically. However, the tunneling rate to the global minimum is {\it extremely} small (See Supplementary Material). 

It is also useful to estimate the strong coupling scale of the model. For this, it is simplest to work in the Stueckelberg formalism, in which one introduces 4 scalar fields $\phi^\ai$ via the replacement $\eta_{\mu\nu}\to \pd_\mu \phi^\ai \pd_\nu \phi^\bi \eta_{\ai\bi}$ in action (\ref{mvmg}) to restore diffeomorphism invariance  (see, {\it e.g.}, \cite{deRham:2014zqa} for details). Around  Minkowski space $g_{\mu\nu}=\eta_{\mu\nu}$ and the trivial $\si$ background $\si=0$, the interaction operators are those of the dRGT model, which are strongly coupled at $\Lambda_3=(m_0^2M_P)^{1/3}$, plus some extra operators that are strongly coupled above $\Lambda_3$ since $m\ll M_P$. The strong coupling scale can be further raised to around $\Lambda_2=(m_0 M_P)^{1/2}$ once the Vainshtein mechanism or equivalently a $\Lambda_2$-type background(which is approximately flat but not of the standard Minkowski form \cite{deRham:2016plk}) is taken into account. On the other hand, in the strong gravity regime, the graviton mass becomes much larger than $m_0$ and, in addition, the effects of $\Lambda_2$-type backgrounds \cite{deRham:2016plk} and curvatures \cite{Aoki:2016vcz} may provide a nontrivial background that further raises the strong coupling scale. So in the strong gravity regime the strong coupling scale can be significantly raised, potentially to around $\Lambda'_2 \sim (m M_P)^{1/2}$. For $m\sim 10/\text{(stellar black hole radius)} \sim \text{km}^{-1}$, we have $\Lambda'_2\sim (10^{-18}\text{km})^{-1}$.
 %%%%%%%%%%%%%%%%%
%%%%%%%%%%%%%%%%%

To quantify how the hairy black hole differs from the GR solution, one can probe its geometry by the geodesic motions of a test object around it. For a static and spherically symmetric solution, we have two Killing vectors $\pd_t$ and $ \pd_\phi$. For the equatorial geodesics ($\theta=\pi/2$), the conserved energy and angular momentum per unit mass are given by $\epsilon = - g_{\mu\nu}(\pd_t)^\mu \ud x^\nu/\ud \li$ and $\ell =- g_{\mu\nu}(\pd_\phi)^\mu \ud x^\nu/\ud \li$ respectively, where $\li$ is the proper time for time-like geodesics and an affine parameter for light-like geodesics. The geodesic equation is given by
\begin{equation}
\frac{E}2 \left(\frac{\ud r}{\ud \li}\right)^2 + V_{\rm eff}=0,~~~V_{\rm eff}\equiv \frac{A}{2} \left(\frac{\ell^2}{ r^2 }+\xi \right) - \frac{\ep^2}{2}   ,
\end{equation}
where $E = A C+B^2$ and $\xi= 1 (0)$ for a massive (massless) particle.   

A few interesting observables can be derived from the geodesic equation. The so-callled innermost stable circular orbit (ISCO) radius $r_{\rm ISCO}$ is determined by $V_{\rm eff}=\ud V_{\rm eff}/\ud r=\ud^2 V_{\rm eff}/\ud r^2=0$ with $\xi=1$. The ISCO frequency is given by $\omega_{\rm ISCO}  =\ud \phi/\ud t =  \sqrt{{ A'}/{2r}}\big|_{{\rm ISCO}}$, which can be measured by fitting the continuum X-ray spectra of black hole accretion disks ({\it e.g.,} \cite{McClintock:2011zq, Bambi:2011jq}) or using the quasi-periodic oscillations in the accretion spectrum \cite{Bambi:2012pa}, as well as accurately probed by eLISA in extreme mass-ratio inspirals (EMRIs) \cite{AmaroSeoane:2007aw}. The maximum redshift of an ISCO photon as measured from far away, $z_{\rm max}$, is from a photon emitted from the backward direction of its motion, which can be probed by the iron-K$\ai$ lines \cite{Fabian:1989ej}. The circular photon orbit $r_{\rm ph}$ is specified by $V_{\rm eff}=\ud V_{\rm eff}/\ud r=0$ with $\xi=0$. One can also define the impact parameter $b_{\rm ph}=\ell/\ep|_{\rm ph}$ at the circular photon orbit, which can be measured by gravitational lensing experiments \cite{Johannsen:2010ru}. Also, the Event Horizon Telescope aims to observe the images around the circular photon orbit of supermassive black holes. In Fig.~\ref{fig:para}, the contours show the fractional deviations of these quantities from their GR counterparts. We find for black holes much heavier than $\mpl^2/m$, the quantities above are essentially the same as those in GR. This is not surprising given that such black holes are Schwarzschild-like unless one gets very close to the horizon.

More sensitive tests of our model come from direct observations of GWs from black hole mergers, 
which can probe the regions further closer to the black hole horizon. A full analysis of the GW waveforms requires a non-perturbative numerical treatment, which is unavailable for massive gravity and its extensions. However, we can still make some forecasts on the waveforms of the early inspiral phase and the ringdown phase using perturbation theory.

In the early insprial phase, we can treat the gravitational field as a perturbation around Minkowski space $ g_{\mu\nu} =   \eta_{\mu\nu} + h_{\mu\nu}$. The linear perturbation theory in Lorentz invariant ghost-free massive gravity is ill-defined within the Vainshtein scale due to the van Dam-Veltman-Zakharov discontinuity. A simple way to circumvent this discontinuity is to replace the graviton mass with the non-Fierz-Pauli form of \cite{Finn:2001qi}, in which case the BD ghost and the longitudinal mode precisely cancel each other \cite{deRham:2016nuf}. In dRGT massive gravity, the Vainshtein 
mechanism screens out the effects of the helicity-0 mode well within the Vainshtein radius. The bounds put on the graviton mass in this non-Fierz-Pauli theory are about the helicity-2 modes and thus applicable to dRGT massive gravity. Defining $\bar{h}_{\mu\nu} \equiv h_{\mu\nu} - \frac{1}{2}\eta_{\mu\nu} h$, the modified perturbative Einstein equation can then be cast as $\mc{E}^{\ri\si}_{\mu\nu}\bar{h}_{\ri\si} + \frac12 V(\si)\bar{h}_{\mu\nu} = \frac{1}{M_P^2} T_{\mu\nu}$, where $\mc{E}^{\ri\si}_{\mu\nu}\bar{h}_{\ri\si}$ is the linearized Einstein tensor and $T_{\mu\nu}$ is the energy momentum tensor of matter. 

The orbit decays very slowly in the early inspiral phase, and therefore we may expect that $V$ changes very slowly with time, $\ud\sqrt{V}/\ud t \ll \omega^2$, where $\omega$ is the frequency of the GWs. Indeed, for black holes of astrophysical interests, we checked that this adiabatic condition, {\it e.g.} $\ud V^{1/2}/\omega^2\ud t < 0.1$, is justified at least for regions down to $3 r_g$, where the GW quadruple formula may cease to be adequate anyway. Therefore, to a good approximation, the GWs emitted at a given orbit can be calculated with the constant $V$ there. Corrections to the GW emission due to a constant graviton mass have been computed in \cite{Finn:2001qi}. For two identical black holes in a circular orbit, the corrections to the GW strain in the frequency domain can be estimated by $h(f)= h_{\rm GR}(f) \left(1+\frac56 \frac{V}{\omega^2}\right)^{-1/2}$, where $f=\omega/2\pi$ and $h_{GR}(f) \propto f^{-7/6}$ is the Fourier transformation of the inspiral waveform predicted by GR.

%%%%%%%%%%%%%%%%%%%%%%%%
\begin{figure}[tbp]
\centering 
\includegraphics[width=0.4\textwidth]{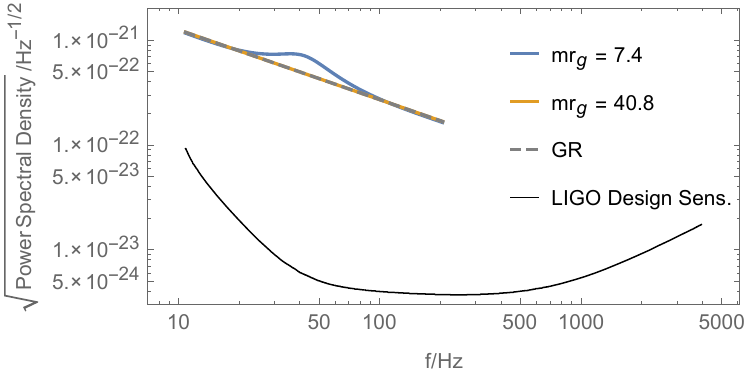}
\caption{Power spectral density of the GW signals sourced by two $30 M_{\odot}$ black holes inspiraling in a circular orbit at a distance of $410 {\rm Mpc}$. The frequencies are cut off at a separation of $3r_g$. The gray dashed line is for GR, while the blue and orange lines are for MVMG with $mr_g = 7.4$ and $mr_g = 40.8$ respectively ($m_\si/m = 10^{-2.25}$). The solid black line shows the design sensitivity of advanced LIGO.}
\label{fig:inspiral}
\end{figure}
%%%%%%%%%%%%%%%%%%%%%%%%

Furthermore, there is also a correction to the waveform due to the varying of the graviton mass as GWs propagate to the observer. This effect can be estimated using a WKB approximation, which leads to an additional factor of $(1-{V}/{\omega^2})^{-1/4}$ correction to the strain (and a correction of $-\frac12 \int_\Delta^\infty \ud r {V}/{\omega^2}$ to the phase, with $\Delta$ being the separation between two black holes when the GWs are emitted). In Fig.~\ref{fig:inspiral}, we plot two examples of the power spectral density of the inspiral waveform, as well as the corresponding GR curve, against the advanced LIGO design sensitivity curve.

The post-merger GWs contain information very near the horizon where the most significant departure from GR occurs. As the characteristic ``sound" of a black hole, the quasi-normal modes (QNMs) are important information encoded in the post-merger waveform. Given the large graviton mass and distinct geometry near the horizon, a quite different QNM spectrum can be expected from that of the Schwarzschild black hole \cite{ZhangZhou}. The early ringdown waveform, however, can still be very similar to the GR waveform,
since the early ringdown waveform is most sensitive to physics close to the circular photon orbit \cite{Cardoso:2016rao}. 

Moreover, as pointed out in \cite{Cardoso:2016rao,Cardoso:2017njb}, dramatic modifications near the horizon typically result in ``echoes'' of the GWs that can be detected by LIGO in the near future. Different from the previous results where exotic matter or quantum effects are invoked, here it is the graviton field itself that undergoes a peculiar change. As a demonstration, we shall consider a test scalar field that couples to $\si$ in a way similar to that of the graviton, and scatter off a wavepacket of such field with initial configurations:
\begin{equation}
\label{wavepaket}
\frac{\pd \Psi_{lm}}{\pd t} (t=0,r) = e^{-\frac{(r_*-r_0)^2}{r_\si^2}}, ~\Psi_{lm}(t=0,r)=0,
\end{equation}
on the black hole background. Here $r_*$ is the tortoise coordinate defined by $dr/dr_*=\sqrt{P Q}$ with $P$ and $1/Q$ being the $tt$ and $rr$ component of the diagonalized background metric respectively. $\Psi_{lm}/r$ is the $lm$ spherical harmonic component of the wavepacket, which satisfies the modified Regge-Wheeler equation
\begin{equation}
\left[-\frac{\pd^2}{\pd t^2}+\frac{\pd^2}{\pd r_*^2} - U_{\rm eff}(r)\right]\Psi_{lm}(t,r)=0,
\end{equation}
with $U_{\rm eff }(r) = P \left( \frac{l(l+1)}{r^2} +\frac12 \frac{\ud Q}{r\ud r} + \frac{Q}{ 2P}\frac{\ud P}{r\ud r} + V\right)$.  
%%%%%%%%%%%%%%%%
\begin{figure}[tbp]
\centering 
\includegraphics[width=0.42\textwidth]{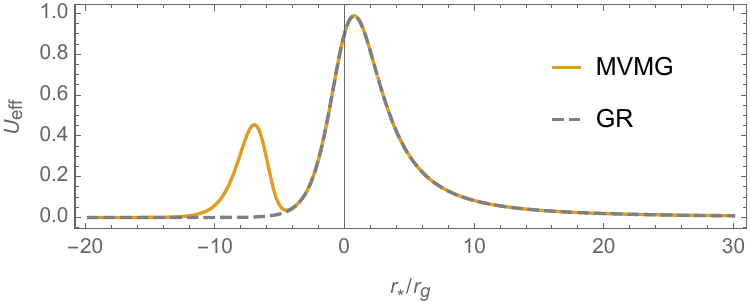}
\includegraphics[width=0.43\textwidth]{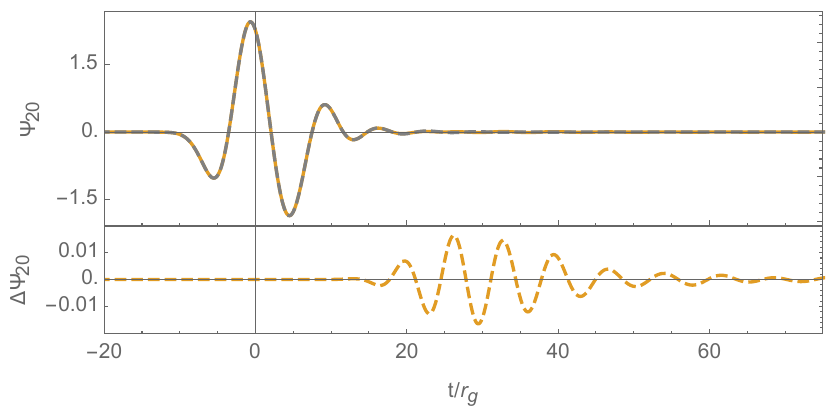}
\caption{The upper panel shows the effective potential $U_{\rm eff}$ with $l=2$ against the tortoise coordinate $r_*$ in GR (gray dashed line) and MVMG (orange solid line). The lower panel shows the waveforms of a test scalar wavepacket scattered in GR (gray dashed line) and MVMG (orange solid line) black hole background. The orange dashed line shows the difference between the GR and MVMG waveforms, which are echoes caused by the extra barrier in MVMG. The wavepacket has the initial configuration~Eq.~(\ref{wavepaket}) with $r_0 = 5 r_g$ and $r_\si=3r_g$. For MVMG, we choose $m_\si = 10^{-1.625}m$ and $mr_g = 648.6$. }
\label{fig:ringdown}
\end{figure}
%%%%%%%%%%%%%%%%%%
An example of the effective potential $U_{\rm eff}$ is shown in Fig.~\ref{fig:ringdown}, where an extra barrier arises due to the enhancement of the graviton mass near the horizon. This results in echoes in the late-time ringdown waveform; See the bottom panel of Fig.~\ref{fig:ringdown}, where the numerical solution of $\Psi_{20}$ is shown for an example. The latency of the echoes depends on the separation between the two barriers \cite{Cardoso:2016oxy}.  In our case, a heavier black hole has an effective potential modified closer to the horizon and thus a further separated double barrier, leading to more prominent echoes. Note that this is opposite of the geodesic deviations, where a lighter black hole deviates more significantly from GR, which will help constrain the model in upcoming experiments.

In summary, we have constructed an extension of dRGT massive gravity where the graviton mass is very small in conventional environments such as in the solar system and other weak gravity regimes, but can be significantly enhanced in the vicinity of black holes. As a result, sizeable observational deviations from GR arise in the strong gravity regime (specially very close to the black hole horizon), which may be examined in the upcoming tests of gravity. Particularly, an extra barrier in the effective potential of black hole perturbations can emerge near the horizon, modifying the GWs from black hole mergers by producing ``echoes'' in the late time ringdown waveform, which will be measured more accurately by LIGO and upcoming GW observations. To our knowledge, the solution in MVMG is the first explicit black hole solution that can lead to ``echoes" in the waveform. Besides, as can be seen in Fig.~\ref{fig:para}, smaller black holes deviate more significantly from GR. Thus, small primordial black holes, if observed, can also provide another good test of this model.

\vskip 20pt
\begin{acknowledgments}

\noindent{\bf Acknowledgments}: We would like to thank Enrico Barausse, Vitor Cardoso, Claudia de Rham, Matthew Johnson, Georgios Pappas and Andrew Tolley for helpful discussions. We also thank Vitor Cardoso for the help with the scattering simulation. JZ is supported by the National Science and Engineering Research Council through a Discovery grant. This research was supported in part by Perimeter Institute for Theoretical Physics. Research at Perimeter Institute is supported by the Government of Canada through the Department of Innovation, Science and Economic Development Canada and by the Province of Ontario through the Ministry of Research, Innovation and Science. SYZ acknowledges support from the starting grant of USTC, the 1000 Young Talent Program of China and the European Union's Horizon 2020 Research Council grant 724659 MassiveCosmo ERC- 2016-COG. 

\end{acknowledgments}

\appendix

\section{Appendix}

In this appendix, we show that the tunneling rate of the $\si=0$ vacuum is extremely small. The tunneling rate per volume can be estimated by $\Gamma/{\cal V} \sim m_\si^4e^{-S_{\text{E}}}$, where $S_{\rm E}$ is the Euclidean action for the bounce solution.  The vacuum not having decayed by now requires $(\Gamma/{\cal V}){\cal T}^4 \lesssim 1$, where ${\cal T} \sim 10^{45}{\rm TeV}^{-1}$ is the age of the universe. This imposes the condition that the minimum bounce action must be $S^{\rm cosmo}_{\rm E} \gtrsim 400 + 4 \ln (m_\si/{\rm TeV})$. On the other hand, our model has $S^{\rm mvmg}_{\rm E} = \mpl^2 \hat{S}^{\rm mvmg}_{\rm E}/(k_3^2 m^2)$, where $\hat{S}^{\rm mvmg}_{\rm E}$ is the re-scaled Euclidian action
\begin{equation}
\hat{S}^{\rm mvmg}_{\rm E} = 2\pi^2\!\! \int_0^{\infty}\!\!\! \rho^3 d\rho \left[ \frac12 \left(\frac{\ud \hat{\sigma}}{\ud \ri}\right)^2 \!\!+ \hat{\sigma}^2- \hat{\sigma}^4 + \kappa_{\si} \hat{\sigma}^6 \right]
\end{equation}
with $\hat{\sigma} = \sqrt{2}k_3 m\si/m_\si$, $\lambda_\si=2\kappa_{\si} k_3^4 m^4/m_\sigma^2$ and $\rho = m_\si\sqrt{(-i t)^2 + r^2}/\sqrt{2}$. By explicitly calculating the bounces for different $\kappa_\si$, one can show that $\hat{S}^{\rm mvmg}_{\rm E}$ monotonically decreases from about $10^4$ to $6.58$ as $\kappa_\si$ decreases from $1/4$ (so that $\si = 0$ is a local minimum) to $0$ (the thick wall limit). $k_3$ is naturally chosen to be order 1 or less, and $M_P^2/m^2$ is a huge ratio since $1/m$ is an astrophysical length scale  in our model.  Therefore, we see that $S^{\rm mvmg}_E$ in our model is much greater than the lower bound of $S^{\rm cosmo}_{\rm E}$, which means the asymptotical vacuum is stable for much far longer than the current age of universe.

\bibliography{references}

\end{document}